# An analytic solution for one-dimensional quantum walks

Ian Fuss[1], Lang White[1], Peter Sherman[2] and Sanjeev Naguleswaran[1]

1. School of Electrical and Electronic Engineering, University of Adelaide, Australia
2. Department of Aerospace Engineering, Iowa University, United States of America

The first general analytic solutions for the one-dimensional walk in position and momentum space are derived. These solutions reveal, among other things, new symmetry features of quantum walk probability densities and further insight into the behaviour of their moments. The analytic expressions for the quantum walk probability distributions provide a means of modelling quantum phenomena that is analogous to that provided by random walks in the classical domain.

## 1. Introduction

### 1.1. Background

There is a growing body of evidence that quantum walks have a similar role in the modelling of quantum phenomena and the development of algorithms for quantum computation as that of random walks in the classical domain [1-5]. This evidence has been accumulated in part through numerical, moment and limit analysis of the discrete one-dimensional quantum walk.

The existence of a general analytic solution for the discrete quantum walk has been an open question thus far. In this paper, such a solution for the position and momentum space wavefunctions of the discrete one-dimensional quantum walk is derived. In contrast to previous analyses of quantum walks that have also utilised the momentum space our analysis provides elementary closed form solutions for both the momentum and position space. These solutions are powerful tools for analysing the properties of quantum walks as well as providing a capability for modelling quantum phenomena.

Quantum walk models have an interesting and important application with regard to whether human decision and reaction time data manifest quantum effects [6, 7]. In this context the linearity of the quantum walk probability distributions, derived in this paper, with respect to their symmetry parameters should enable them to be estimated from experimental data via simple techniques, while the isolation of the temporal evolution parameter enables its estimation via a one-dimensional search.

Previous work that has been motivated by the development of algorithms for quantum computing has sought insight for this by choosing paradigmatic examples of quantum walks, such as the Hadamard walk, and argued that these are typical of quantum walks as a whole [1, 3]. The work of this paper in developing simple explicit analytic forms of the general wave functions and hence probability distributions, that are closely connected to the underlying parameters, contrasts to this paradigmatic approach. The results obtained raise questions about how typical the Hadamard walk is of the spectrum of discrete quantum walks.

Hence while it was not the main motivation in seeking these general solutions it is possible that they will provide a better basis for developing quantum algorithms than the approaches used thus far. For example the isolation of the temporal evolution parameter in the walk from the other two symmetry parameters may assist in developing optimisation strategies for lattice searches to locate particular values of a parameter [9] or to determine a satisfactory plan [10].

### 1.2. Structure

In the next section the quantum walk dynamic equations in position space are stated and used to obtain those for the momentum space. Then in section 3 a generalised de Moivre principal is used to derive the momentum-space time-evolution operator in terms of type II Chebyshev polynomials.

Analytic expressions for the momentum space $\phi(p,t)$ and position space $\psi(x,t)$ quantum walk wavefunctions are obtained using this time-evolution operator in section 4. The position space wavefunctions are left in a partially abstract form in this section to assist in providing a general analysis of the symmetry properties of the quantum walk position space probability densities $\rho(x,t)$ in section 5. The key result of section 5 is that the quantum walk probability distributions have the special form of a sum of even and odd functions. These odd functions are multiplied by the two quantum walk parameters, previously referred to as symmetry parameters.

An explicit analytic representation of the position space wave functions is presented in section 6. This makes use of a new type of one-dimensional function whose properties are partially analysed to support the subsequent quantum walk analysis. In section 7 these functions are used to derive analytic expressions for the moments of the walks. In section 8 the results of the moment analysis are used to analyse the walks and provide linkages to other results in the literature. The paper finishes with a summary and the mention of an open area of research.

## 2. Quantum Walk Dynamic Equations

For a given $\psi(x,0) = \begin{pmatrix} c_0 \\ c_1 \end{pmatrix} \delta_{x,0}$ we consider the evolution of a quantum state $\psi(x,t) \in C^2$ for discrete times $t \geq 0$ on a line $x \in Z$. [10, 11] The dynamics of the state evolve according to the difference equations

$$\psi_0(x,t) = e^{ik}[a\psi_0(x-1,t-1) + b\psi_1(x-1,t-1)]$$

$$\psi_1(x,t) = e^{ik}[-b^*\psi_0(x+1,t-1) + a^*\psi_1(x+1,t-1)] \qquad (1)$$

where $|a|^2 + |b|^2 = 1$, $k \in R$ and $\psi_0$ and $\psi_1$ are the components of the spinor $\psi$.

We note that these difference equations define a linear system and hence follow Nayak and Vishwanath in exploiting the spatial homogeneity of the quantum walks by applying the Fourier transform [1]

$$\phi(p,t) = \sum_{x=-\infty}^{\infty} \psi(x,t) e^{-ixp} \quad (2)$$

to equation 1 to give

$$\phi_0(p,t) = e^{ik}\left[ae^{-ip}\phi_0(p,t-1) + be^{-ip}\phi_1(p,t-1)\right]$$

$$\phi_1(p,t) = e^{ik}\left[-b^*e^{ip}\phi_0(p,t-1) + a^*e^{ip}\phi_1(p,t-1)\right]. \quad (3)$$

We note that with a choice of units so that $\hbar = 1$ and an appropriate normalisation procedure such as requiring periodic boundary conditions then the spinor

$$\phi(p,t) = \begin{pmatrix} \phi_0(p,t) \\ \phi_1(p,t) \end{pmatrix} \quad (4)$$

can be interpreted as a momentum space wave-function for the quantum walk with $p \in (-\pi, \pi]$ [12].

Using this wave function equation 3 can be rewritten as the matrix equation

$$\phi(p,t) = e^{ik} S(p) \phi(p,t-1) \quad (5)$$

where

$$S(p) = \begin{pmatrix} ae^{-ip} & be^{-ip} \\ -b^*e^{ip} & a^*e^{ip} \end{pmatrix} \quad (6)$$

is a unimodular matrix.

By applying the Fourier transform to the initial condition $\psi(x,0)$ we can obtain the equivalent initial condition in the momentum space

$$\begin{pmatrix} \phi_0(p,0) \\ \phi_1(p,0) \end{pmatrix} = \begin{pmatrix} \psi_0(0,0) \\ \psi_1(0,0) \end{pmatrix}. \quad (7)$$

We can obtain the momentum space wavefunction at an arbitrary time with this initial condition by iterating equation 5 to give

$$\phi(p,t) = e^{itk} S^t(p) \phi(p,0). \quad (8)$$

This equation permits the interpretation of

$$T(t_1 - t_0) = e^{i(t_1-t_0)k} S^{(t_1-t_0)}(p) \tag{9}$$

as the evolution operator in the momentum space. This 2 by 2 matrix polynomial provides a simpler analytic base than its equivalent the infinite dimensional constant matrix in the position space.

## 3. The Evolution Operator

Other authors have analysed the momentum space wave functions using an eigenvalue decomposition of this evolution operator [1, 3]. We use an alternate approach that has the advantage of giving the general evolution operator in a simple analytic form and hence general elementary closed forms for the momentum and position space wavefunctions.

The unimodular matrix $S(p)$ can be written in an exponential form as

$$S(p) = Exp(i\theta(p)\mathbf{c}(p).\boldsymbol{\sigma}) = \cos(\theta(p))I + i\sin(\theta(p))\mathbf{c}(p).\boldsymbol{\sigma} \tag{10}$$

where $\theta$ and $c$ are real functions of $p$ and the matrix vector $\boldsymbol{\sigma}$ has Pauli matrix components [13, 14]

$$\sigma_1 = \begin{bmatrix} 0 & 1 \\ 1 & 0 \end{bmatrix}, \sigma_2 = \begin{bmatrix} 0 & -i \\ i & 0 \end{bmatrix} \text{ and } \sigma_3 = \begin{bmatrix} 1 & 0 \\ 0 & -1 \end{bmatrix} \tag{11}$$

and $I$ is the identity matrix.

This exponential form allows us to can write the powers of $S(p)$ as[1]

$$S^t(p) = Exp(it\theta(p)\mathbf{c}(p).\boldsymbol{\sigma}) = \cos(t\theta(p))I + i\sin(t\theta(p))\mathbf{c}(p).\boldsymbol{\sigma}. \tag{12}$$

The trigonometric expressions in this equation can be written in terms of the Chebyshev polynomials $T_n$ and $U_n$ as [15]

$$\cos(t\theta) = T_t(\cos(\theta))$$

and

$$\sin(t\theta) = U_{t-1}(\cos(\theta))\sin(\theta). \tag{13}$$

Substituting these expressions into equation 12 gives

$$S^t(p) = T_t(\cos(\theta(p)))I + U_{t-1}(\cos(\theta(p)))i\sin(\theta(p))\mathbf{c}(p).\boldsymbol{\sigma}. \tag{14}$$

---

[1] It is interesting to note that this form of the time evolution operator implies that the momentum space representation of the Hamiltonian for the quantum walk is $H = \theta(p)\mathbf{c}(p).\boldsymbol{\sigma}$.

This equation can be written in terms of the type II Chebyshev polynomials alone by using the relation

$$T_n(x) = U_n(x) - x U_{n-1}(x) \qquad (15)$$

to give

$$S^t(p) = U_t(\cos(\theta(p)))I - U_{t-1}(\cos(\theta(p)))[\cos(-\theta(p))I + i\sin(-\theta(p))\mathbf{c}(p)\cdot\boldsymbol{\sigma}] \qquad (16)$$

where we have used the even property of the cosine function and the odd property of the sine function.

By comparing the expression within the square brackets with equation 10 we see it is equivalent to an exponentiation of the inverse of the one step time evolution matrix $S^{-1}(p)$ and thus write

$$S^t(p) = U_t(\cos(\theta(p)))I - U_{t-1}(\cos(\theta(p)))S^{-1}(p). \qquad (17)$$

By substituting this expression into equation 9 we obtain the time evolution operator as

$$T(t,0) = e^{itk}\left[U_t(\cos(\theta(p)))I - U_{t-1}(\cos(\theta(p)))S^{-1}(p)\right]. \qquad (18)$$

We can determine the function $\cos(\theta(p))$ in terms of the components of $S(p)$ by defining the inner product

$$(A,B) = \frac{1}{2}Tr(AB) \qquad (19)$$

on the vector space of two by two unitary matrices and hence obtain an inner product space with $\{I, \sigma_1, \sigma_2, \sigma_3\}$ as an ortho-normal basis.

We note that the coefficient of $I$ on the right had side of equation 10 is $\cos(\theta(p))$ and the coefficient of $I$ on the left hand side is

$$(I, S(p)) = |a|\cos(p-d) \qquad (20)$$

where $|a|$ is the absolute value of $a$ and $d$ its argument, that is

$$a = |a|e^{id}. \qquad (21)$$

Thus using the equality of coefficients and equation 20 we can write the time evolution operator in equation 18 as

$$T(t,0) = e^{itk}\left\{U_t(|a|\cos(p-d))I - U_{t-1}(|a|\cos(p-d))S^{-1}(p)\right\}. \qquad (22)$$

# 4. Wave Functions

## 4.1. Momentum Space

The momentum space wave functions are now easily obtained from equations 8, 9 and 22 as

$$\phi(p,t) = e^{itk}\{U_t(|a|\cos(p-d))I - U_{t-1}(|a|\cos(p-d))S^{-1}(p)\}\phi(p,0) \quad (23)$$

or using

$$\phi(p,-1) = S^{-1}(p)\phi(p,0), \quad (24)$$

as

$$\phi(p,t) = e^{itk}U_t(|a|\cos(p-d))\phi(p,0) - e^{ikt}U_{t-1}(|a|\cos(p-d))\phi(p,-1). \quad (25)$$

We can write this more explicitly in terms of the spinor components as

$$\phi_0(p,t) = e^{itk}\left[U_t(|a|\cos(p-d))c_0 - U_{t-1}(|a|\cos(p-d))(|a|e^{i(p-d)}c_0 - \beta e^{-i(p-d)}c_1)\right] \quad (26a)$$

$$\phi_1(p,t) = e^{itk}\left[U_t(|a|\cos(p-d))c_1 - U_{t-1}(|a|\cos(p-d))(\beta^* e^{i(p-d)}c_0 + |a|e^{-i(p-d)}c_1)\right] \quad (26b)$$

where the latter terms are derived from the expansion of equation 24 as

$$\phi(p,-1) = \begin{pmatrix} a^* e^{ip} & -be^{-ip} \\ b^* e^{ip} & ae^{-ip} \end{pmatrix}\begin{pmatrix} c_0 \\ c_1 \end{pmatrix} \quad (27)$$

and defining

$$\beta = be^{-id}. \quad (28)$$

## 4.2. Position Space

The inverse Fourier transform of equation 26 gives the position space wave functions

$$\psi_0(x,t) = e^{i(xd+tk)}\left[c_0(u_t(|a|:x) - |a|u_{t-1}(|a|:x+1)) + \beta c_1 u_{t-1}(|a|:x-1)\right] \quad (29)$$

$$\psi_1(x,t) = e^{i(xd+tk)}\left[c_1(u_t(|a|:x) - |a|u_{t-1}(|a|:x-1)) - \beta^* c_0 u_{t-1}(|a|:x+1)\beta^* c_0\right] \quad (30)$$

where have defined

$$u_t(|a|:x) = \frac{1}{2\pi}\int_{-\pi}^{\pi} U_t(|a|\cos(p))e^{ixp}\,dp \quad (31)$$

as the inverse Fourier transform of the augmented Type II Chebyshev polynomials.

We will show in section 6 that the functions $u_t(|a|:x)$ can be expressed in a simple closed form. At present we note that they are real even functions as they are the inverse Fourier transform of real even functions.

An inspection of equations 29 and 30 leads us to define the function

$$f_t(|a|:x) = u_t(|a|:x) - |a|u_{t-1}(|a|:x+1) \qquad (32)$$

and hence write the position space wavefunctions as

$$\psi_0(x,t) = e^{i(xd+tk)}[c_0 f_t(|a|:x) + \beta c_1 u_{t-1}(|a|:x-1)] \qquad (33a)$$

$$\psi_1(x,t) = e^{i(xd+tk)}[c_1 f_t(|a|:-x) - \beta^* c_0 u_{t-1}(|a|:-x-1)]. \qquad (33b)$$

The expressions in equations 29, 30 and 33 show clearly that the detailed temporal dynamics of the quantum walk depend dominantly on the single real parameter $|a|$ via the functions $u_t(|a|:x)$.

## 5. Probability Densities and Moments

### 5.1. Probability Densities

The quantum walk probability densities for each of the spinor components can be written by using equation 33 as

$$\rho_0(x,t) = |c_0|^2 f_t^2(|a|:x) + 2|bc_0 c_1|\cos(\delta)f_t(|a|:x)u_{t-1}(|a|:x-1) \qquad (34a)$$
$$+ |b|^2|c_1|^2 u_{t-1}^2(|a|:x-1)$$

$$\rho_1(x,t) = f_t^2(|a|:-x)|c_1|^2 - 2|bc_0 c_1|\cos(\delta)f_t(|a|:-x)u_{t-1}(|a|:-x-1) \qquad (34b)$$
$$+ |b|^2|c_0|^2 u_{t-1}^2(|a|:-x-1)$$

where

$$\delta = \arg(a) - \arg(b) + \arg(c_0) - \arg(c_1). \qquad (35)$$

Thus it can be seen explicitly that the one dimensional quantum walk is an interferometric system with only three effective real parameters that can be chosen as $|a|, |c_0|$ and $\delta$. We note that $|b|^2 = 1 - |a|^2$ and $|c_1|^2 = 1 - |c_0|^2$.

By expanding the functions $f_t^2$ and $f_t$ we are able to write the quantum walk probability distributions as the sum of an even

$$\rho_{even}^{|a|}(x,t) = u_t^2(|a|:x) + \frac{1}{2}\left[u_{t-1}^2(|a|:x-1) + u_{t-1}^2(|a|:x+1)\right] - |a|u_t(|a|:x)\left[u_{t-1}(|a|:x-1) + u_{t-1}(|a|:x+1)\right], \quad (36a)$$

and odd component

$$\rho_{odd}^{|a|,\alpha,\nu}(x,t) = \nu\left[u_{t-1}^2(|a|:x-1) - u_{t-1}^2(|a|:x+1)\right] + \quad (36b)$$
$$(2|a|\nu - \alpha)u_t(|a|:x)\{u_{t-1}(|a|:x-1) - u_{t-1}(|a|:x+1)\}$$

where

$$\nu = |c_0|^2 - \frac{1}{2} = \frac{1}{2}\left(|c_0|^2 - |c_1|^2\right) \quad (37)$$

and

$$\alpha = 2|bc_0 c_1|\cos(\delta). \quad (38)$$

We note that the sum of even and odd functions is a special form for an asymmetric function. Examples of the even and odd components are plotted in figures 1 and 2.

Further we note that the quantum walk probability distributions are linear in the parameters $\nu$ and $\alpha$ and that these affect only their symmetry. These parameters and hence the odd component of the distribution are zero when

$$|c_0| = \frac{1}{\sqrt{2}} \quad (39)$$

and

$$\cos(\delta) = 0. \quad (40)$$

Thus these are sufficient conditions for a quantum walk to be even.

## 5.2. Moments 1

Since the quantum walk probability densities can be decomposed exactly into the sum of an even and odd component then the even moments of the quantum walk

$$\left\langle x^{2n}\right\rangle_t^{|a|} = \sum_{x=-\infty}^{\infty} x^{2n}\rho(x,t) = \sum_{x=-\infty}^{\infty} x^{2n}\rho_{even}^{|a|}(x,t), \text{ where } n \in N, \quad (41)$$

depend only on the even component of the probability distribution, equation 36a and hence only on the parameter $|a|$.

Similarly the odd moments of the quantum walk

$$\left\langle x^{2n+1}\right\rangle_t^{|a|,\alpha,\nu} = \sum_{x=-\infty}^{\infty} x^{2n+1}\rho(x,t) = \sum_{x=-\infty}^{\infty} x^{2n+1}\rho_{odd}^{|a|,\alpha,\nu}(x,t), \text{ where } n \in N \quad (42)$$

depend only on the odd component of the probability distribution equation 36b. The form of the odd probability distribution leads naturally to the further decomposition

$$\rho_{odd}^{|a|,\alpha,\nu}(x,t) = (2\nu - \alpha)\tilde{\rho}_{mi}^{|a|}(x,t) - \nu\tilde{\rho}_{sq}^{|a|}(x,t). \tag{43}$$

with components
$$\tilde{\rho}_{sq}^{|a|}(x,t) = u_{t-1}^2(|a|:x-1) - u_{t-1}^2(|a|:x+1) \tag{44}$$

and
$$\tilde{\rho}_{mi}^{|a|}(x,t) = u_t(|a|:x)\{\hat{U}_{t-1}(|a|:x-1) - u_{t-1}(|a|:x+1)\}. \tag{45}$$

Thus we can write the odd moments as

$$\left\langle x^{2n+1}\right\rangle_t^{|a|,\alpha,\nu} = (2|a|\nu - \alpha)\sum_{x=-\infty}^{\infty} x^{2n+1}\tilde{\rho}_{mi}^{|a|}(x,t) - \nu\sum_{x=-\infty}^{\infty} x^{2n+1}\tilde{\rho}_{sq}^{|a|}(x,t). \tag{46}$$

### *5.3.    Calculations*

The even probability distribution equation 36a can be simplified to

$$\rho_{even}^{|a|}(x,t) = \frac{1}{2}\left[u_{t-1}^2(|a|:x-1) + u_{t-1}^2(|a|:x+1)\right] - u_t(|a|:x)u_{t-2}(|a|:x). \tag{47}$$

by using the recursion

$$u_{t+1}(|a|:x) = |a|u_t(|a|:x+1) + |a|u_t(|a|:x-1) - u_{t-1}(|a|:x) \tag{48}$$

which is derived in appendix A.

In order to provide a comparison with other elements of the literature equations 47 and 48 were used to calculate the probability distribution in figure 1. [1, 3 and 4]

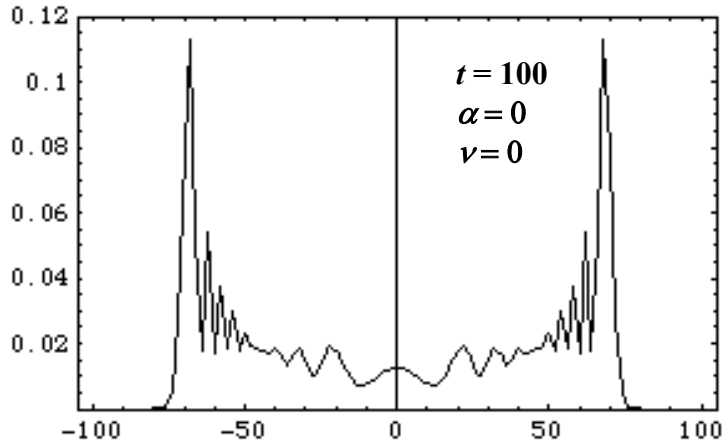

Figure 1. The even Hadamard walk probability distribution at $t = 100$

The odd component of the probability distribution for $\alpha = 0$ is

$$\rho_{odd}^{|a|,\alpha,\nu}(x,t) = \nu\left(2\widetilde{\rho}_{mi}^{|a|}(x,t) - \widetilde{\rho}_{sq}^{|a|}(x,t)\right), \quad (49)$$

where equation 43 was used.

This is plotted for the Hadamard walk at $t = 100$ and $\nu = 1/2$ in Figure 2.

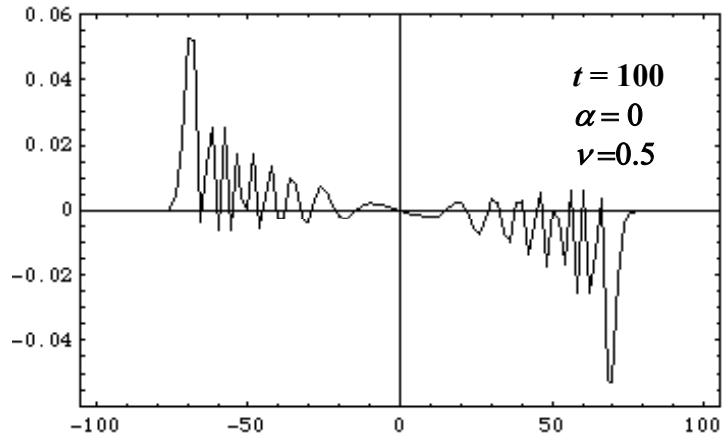

Figure 2. An odd component of the Hadamard walk Probability distribution at $t = 100$

By varying either $\alpha$ or $\nu$ the symmetry of the quantum walk probability distribution is changed by mixing differing amounts of the anti-symmetric component with the symmetric. Figure 3 shows this effect for $\nu$. A similar but less significant effect results from varying $\alpha$.

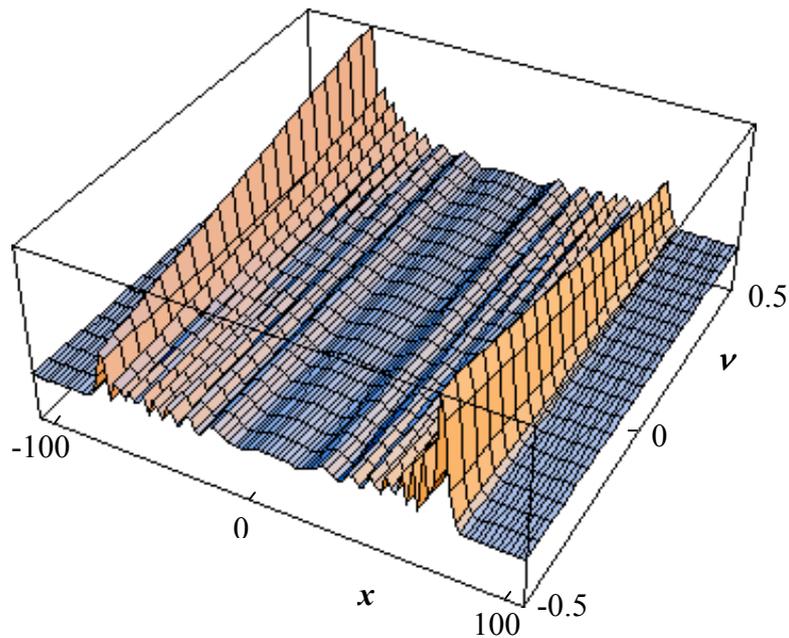

Figure 3. Change in the symmetry of the Hadamard walk with $\nu$ at $t = 100$.

## 6. Foundation Functions

An analysis of the foundation functions $u_t(|a|:x)$ helps develop tools to analyse quantum walks such as algebraic expressions for the moments in terms of the quantum walk parameters. It also helps develop insight into the temporal behaviour of quantum walks.

We start by considering the foundation functions role in building up the various distributions associated with the quantum walks. Figure 4 below shows the function $u_{99}\left(\frac{1}{\sqrt{2}}:x\right)$.

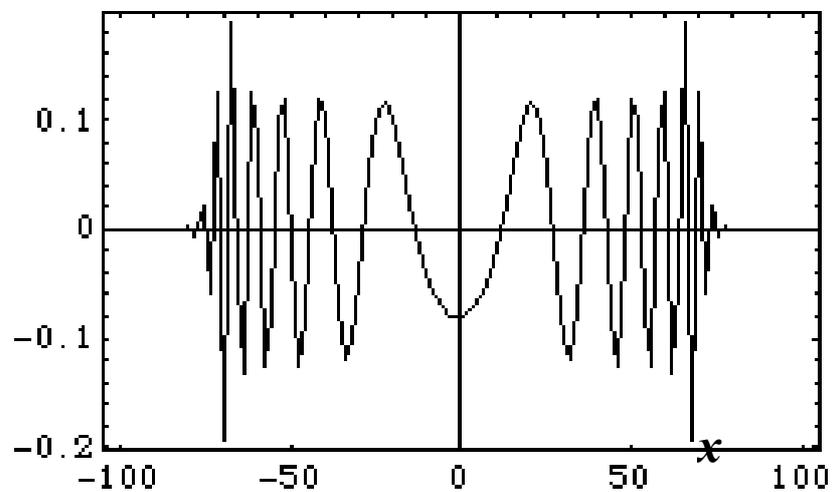

Figure 4. The Foundation Function $u_{99}\left(\frac{1}{\sqrt{2}}:x\right)$

This function provides the major elements of the even component of the Hadamard walk at time $t=100$, that was illustrated in figure 1. The dominant features of this component can be seen in the square of this function that is presented in figure 5.

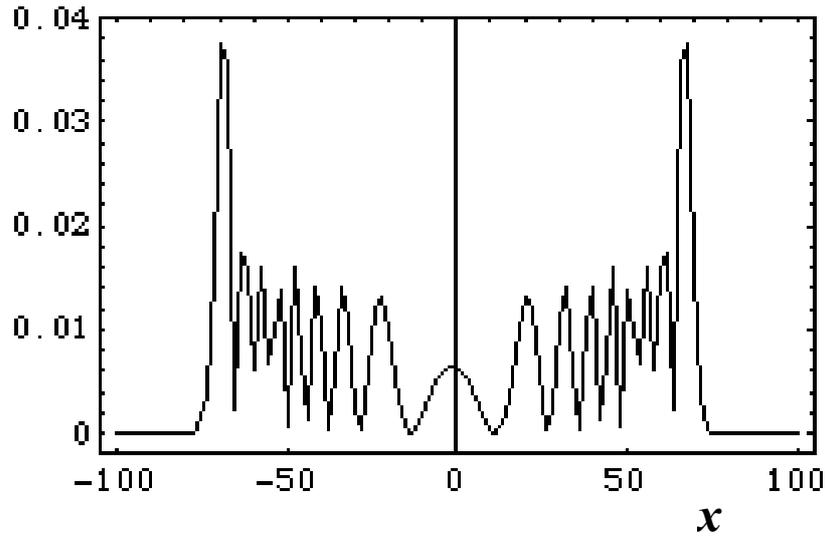

Figure 5. The square $\left| u_{99}\left(\frac{1}{\sqrt{2}} : x\right) \right|^2$

The even component of the quantum walk probability distribution is

$$\rho_{even}^{|a|}(x,t) = \frac{1}{2}\left[u_{t-1}^2(|a| : x-1) + u_{t-1}^2(|a| : x+1)\right] - u_t(|a| : x)u_{t-2}(|a| : x). \quad (47)$$

The first component of this expression is an offset and addition of the square of the foundation function at *t*-1. Such an operation is well known in numerical analysis and signal processing to have the effect of smoothing the function operated on. This effect can be seen in figure 6.

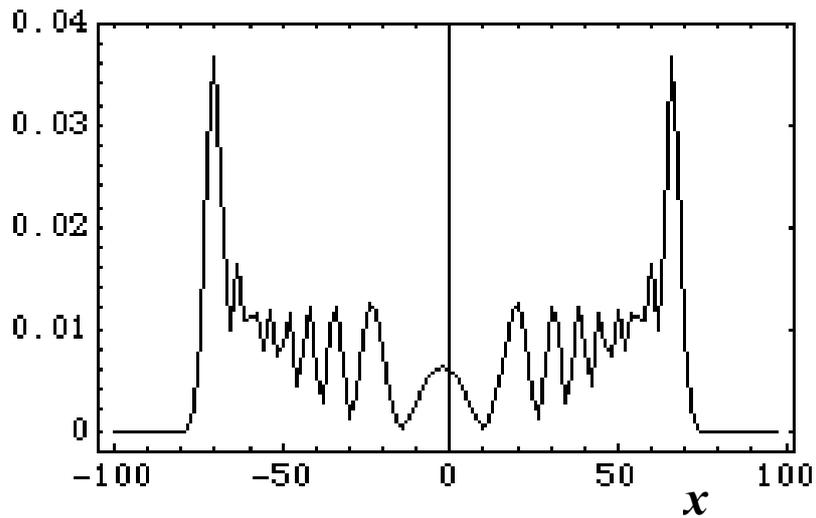

Figure 6. The smoothed square $\left| u_{99}\left(\frac{1}{\sqrt{2}} : x\right) \right|^2$

While the resemblance between the smoothed square of the foundation function and the even component of the quantum walk illustrated in figure 1 is striking we note that the even component is smoother near the origin. This extra smoothing is the result of adding the final term in equation 47.

## 7.1 Polynomial Expressions for the Foundation Functions

We now proceed to develop polynomial expression for these foundation functions by noting that Type II Chebyshev Polynomials can be as expressed as the power series [15]

$$U_t(y) = \sum_{m=0}^{[t/2]} \begin{bmatrix} t \\ m \end{bmatrix} (2y)^{t-2m} \quad (50)$$

with $[t/2]$ indicating the first integer below $t/2$ and

$$\begin{bmatrix} t \\ m \end{bmatrix} = (-1)^m \binom{t-m}{m} \quad (51)$$

where $\binom{l}{k}$ is the binomial coefficient.

By substituting 
$$2y = |a|(e^{ip} + e^{-ip}) \quad (52)$$

in equation 50 and expanding the powers using the binomial theorem

we obtain the expression

$$U_t(|a|\cos(p)) = \sum_{m=0}^{[t/2]} \begin{bmatrix} t \\ m \end{bmatrix} \sum_{k=0}^{t-2m} \binom{t-2m}{k} |a|^{t-2m} e^{-ip(t-2(m+k))} . \quad (53)$$

We would like to put this in the form

$$U_t(|a|\cos(p)) = \sum_{j=0}^{t} P_{t-2j}^t(|a|) e^{-i(t-2j)p} \quad (54)$$

with inverse Fourier transform

$$u_t(|a|:x) = \sum_{j=0}^{t} P_{t-2j}^t(|a|) \delta_{x,t-2j} \quad (55)$$

where the $P_{t-2j}^t(|a|)$ are polynomials in $|a|$.

Hence we change variables in the second summation to $j = k + m$ and use the properties $\binom{l}{k} = 0$ if $k < 0$ or $k > l$ to remove the $m$ dependence from the sum and hence obtain

$$U_t(|a|\cos(p)) = \sum_{m=0}^{[t/2]} \binom{t}{m} \sum_{j=0}^{t} \binom{t-2m}{j-m} |a|^{t-2m} e^{-ip(t-2j)}. \tag{56}$$

Reordering the summations gives

$$U_t(|a|\cos(p)) = \sum_{j=0}^{t} \sum_{m=0}^{[t/2]} \binom{t}{m} \binom{t-2m}{j-m} |a|^{t-2m} e^{-ip(t-2j)} \tag{57}$$

and then comparing this with equation 54 obtain

$$P_{t-2j}^t(|a|) = \sum_{m=0}^{[t/2]} \binom{t}{m} \binom{t-2m}{j-m} |a|^{t-2m}. \tag{58}$$

We note the following properties of the polynomials: They obey the recursion

$$P_k^{t+1}(|a|) = |a| P_{k+1}^t(|a|) + |a| P_{k-1}^t(|a|) - P_k^{t-1}(|a|). \tag{59}$$

This is derived in appendix A.

The even symmetry of $\hat{U}_t(|a|:x)$ with respect to $x$ implies

$$P_k^t(|a|) = P_{-k}^t(|a|). \tag{60}$$

For $j = 0$ the summation in equation (56) truncates at $m = 0$ and thus

$$P_t^t(|a|) = |a|^t, \tag{61}$$

and for $j = 1$ the summation in equation (49) truncates at $m = 1$ and

$$P_{t-2}^t(|a|) = t|a|^t - (t-1)|a|^{t-2} \tag{62}$$

where we have used $\binom{t}{1} = t$.

The formulas in equations 60 to 62 give us the polynomials up to $t = 3$.

For $t = 4$ they are augmented by

$$P_0^4(|a|) = 30|a|^4 - 12|a|^2 + 1 \tag{63}$$

for $t = 5$

$$P_1^5(|a|) = 10|a|^5 - 12|a|^3 + 3|a| \tag{64}$$

and for $t = 6$

$$P_0^6(|a|) = 20|a|^6 - 30|a|^4 + 12|a|^2 - 1 \tag{65}$$

$$P_2^6(|a|) = 15|a|^6 - 20|a|^4 + 6|a|^2 . \tag{66}$$

Finally when $|a| = 1$ then $\hat{U}_t(1:x)$ is the inverse Fourier transform of a Chebyshev polynomial hence

$$P_j^t(1) = 1 \text{ if } |j| \le t \tag{67}$$

and zero otherwise.

This expression is useful for checking the general expressions for the polynomials and quantities derived from them such as the moments of the quantum walk probability distribution functions.

## 7. Moments 2

Analytic expressions for the moments of one-dimensional quantum walks are calculated in the following section by using the foundation polynomials. This allows direct comparison with the moments of experimental data and also connects with the literature on quantum walks that is moment based [1 and 2].

### 7.1 Even Moments

In this subsection expressions for the normalisation and second moment of the quantum walk are calculated using equation 41.

#### 7.1.1 Normalisation

The normalisation requirement is

$$1 = \sum_{x=-\infty}^{\infty} \rho(x,t) = \sum_{x=-\infty}^{\infty} \rho_{even}^{|a|}(x,t). \tag{68}$$

By using equation 47 and making the appropriate changes in summation variables this can be seen to be equivalent to

$$1 = \sum_{x=-\infty}^{\infty} u_{t-1}^2(|a|:x) - \sum_{x=-\infty}^{\infty} u_t(|a|:x) u_{t-2}(|a|:x) \tag{69}$$

which leads via equation 55 to the polynomial relation

$$\sum_{j=0}^{t-1} \left[P_{t-2j-1}^{t-1}(|a|)\right]^2 = 1 + \sum_{j=0}^{t-2} P_{t-2j-2}^{t}(|a|) P_{t-2j-2}^{t-2}(|a|). \tag{70}$$

This equation can be seen to be trivially true in the case $|a|=1$ by using equation 67.

### 7.1.2 Second Moment

The equation

$$\langle x^2 \rangle_t^{|a|} = \sum_{x=-\infty}^{\infty} x^2 \rho_{even}^{|a|}(x,t) \tag{71}$$

for the second moment can be transformed by using equation 47 into

$$\langle x^2 \rangle_t^{|a|} = \sum_{x=-\infty}^{\infty} x^2 u_{t-1}^2(|a|:x) - \sum_{x=-\infty}^{\infty} x^2 u_t(|a|:x) u_{t-2}(|a|:x) + \sum_{x=-\infty}^{\infty} u_{t-1}^2(|a|:x). \tag{72}$$

This equation can be used with the relation between the foundation function and polynomials given in equation 55 to calculate analytic expressions for the second moments such as

$$\langle x^2 \rangle_1^{|a|} = 1 \tag{73}$$

$$\langle x^2 \rangle_2^{|a|} = 4|a|^2 \tag{74}$$

$$\langle x^2 \rangle_3^{|a|} = 8|a|^4 + 1 \tag{75}$$

$$\langle x^2 \rangle_4^{|a|} = 24|a|^6 - 24|a|^4 + 16|a|^2 \tag{76}$$

$$\langle x^2 \rangle_5^{|a|} = 80|a|^8 - 128|a|^6 + 72|a|^4 + 1 \tag{77}$$

and

$$\langle x^2 \rangle_6^{|a|} = 280|a|^{10} - 600|a|^8 + 464|a|^6 - 144|a|^4 + 36|a|^2. \tag{78}$$

In the special case of $|a|=1$ by using equation 67 it can be shown that[2]

$$\langle x^2 \rangle_t^1 = t^2. \qquad (79)$$

Nayak and Vishwanath [1] have shown that in the long time limit the second moment of the Hadamard walk, with $|a|=\frac{1}{\sqrt{2}}$, also increases quadratically with time

$$\langle x^2 \rangle_t^{\frac{1}{\sqrt{2}}} \propto t^2. \qquad (80)$$

Hence in the light of this and equation 79 we investigate the temporal dependence of the normalised second moments

$$M_t(|a|) = \frac{\langle x^2 \rangle_t^{|a|}}{t^2}. \qquad (81)$$

In figure 4 $M_t(|a|)$ is plotted for $t \in \{1,2,3,4,5,6\}$ with odd times as dashed lines and even as full lines. Even for these small times the normalised second moments appear to be converging towards a limit with a monotonic increase of the slope of approach towards the fixed point at $|a|=1$ as a function of time. We further note that the value of the odd time moments $\langle x^2 \rangle_{odd(t)}^0 = 1$ remains fixed and thus the normalised moments $M_{odd(t)}(0) = \frac{1}{t^2}$ converge to zero in the long time limit.

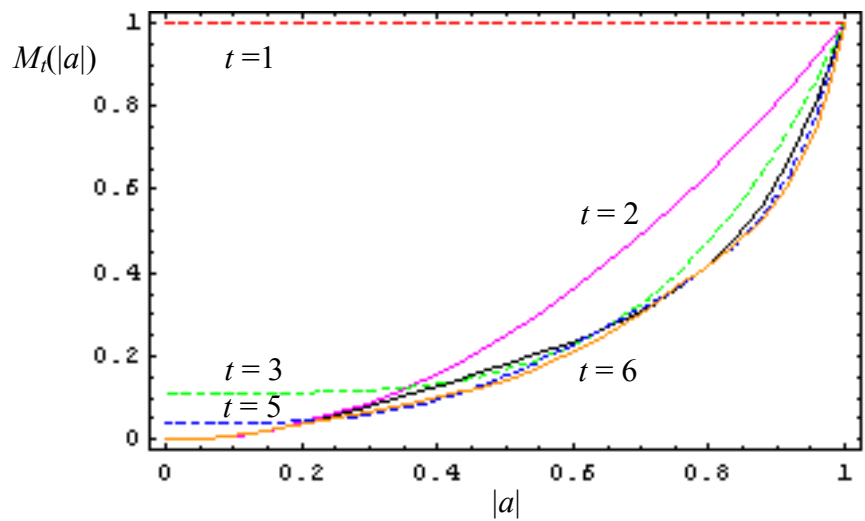

Figure 4. Normalised Second Moments of the Quantum Walk

---

[2] We note that this expression provides a useful check of the more general moment expressions.

In concluding this section we note with foresight that the oscillation in time between zero and one of the second moment at $|a|=0$ is consistent with an oscillation of the first moment at this point.

## 7.2 Odd Moments

By substituting from equation 55 into the expression for the odd moments given in equation 46 we obtain an expression for the odd moments in terms of the foundation polynomials

$$\langle x^{2n+1} \rangle_t^{|a|,\alpha,\nu} = (4|a|\nu - 2\alpha)\sum_{j=0}^{t-1} P_{t-2j}^t(|a|)P_{t-(2j+1)}^{t-1}(|a|)(t-2j)^{2n+1} \quad (82)$$

$$-2\nu\sum_{j=0}^{t-1}\left[P_{t-(2j+1)}^{t-1}(|a|)\right]^2(t-2j)^{2n+1}.$$

### 7.2.1 First Moment

Hence the first moment is

$$\langle x \rangle_t^{|a|,\alpha,\nu} = (4|a|\nu - 2\alpha)\sum_{j=0}^{t-1} P_{t-2j}^t(|a|)P_{t-(2j+1)}^{t-1}(|a|)(t-2j) - 2\nu\sum_{j=0}^{t-1}\left[P_{t-(2j+1)}^{t-1}(|a|)\right]^2(t-2j). \quad (83)$$

At the time $t=1$ the first moment is

$$\langle x \rangle_1^{|a|,\alpha,\nu} = \nu\left[4|a|^2 - 2\right] - 2\alpha|a| \quad (84)$$

that can be written in terms of the parameters $|c_0|$ and $\delta$ as

$$\langle x \rangle_1^{|a|,|c_0|,\delta} = (2|c_0|^2 - 1)(2|a|^2 - 1) - 4|a|\sqrt{1-|a|^2}\,|c_0|\sqrt{1-|c_0|^2}\cos(\delta), \quad (85)$$

by using equation 84, 37 and 38.

We note the symmetry between $|a|$ and $|c_0|$ in this expression and that if $\cos(\delta)=0$ that the moment is quadratic in both of these parameters.

The first moments from $t=2$ to $t=5$ are given below

$$\langle x \rangle_2^{|a|,\alpha,\nu} = \nu\left[8|a|^4 - 4|a|^2\right] - \alpha 4|a|^3 \quad (88)$$

$$\langle x \rangle_3^{|a|,\alpha,\nu} = \nu\left[24|a|^6 - 32|a|^4 + 16|a|^2 - 2\right] - \alpha\left[12|a|^5 - 10|a|^3 + 4|a|\right] \quad (89)$$

$$\langle x \rangle_4^{|a|,\alpha,\nu} = \nu\left[80|a|^8 - 152|a|^6 + 96|a|^4 - 16|a|^2\right] - \alpha\left[40|a|^7 - 56|a|^5 + 24|a|^3\right] \quad (90)$$

$$\langle x \rangle_5^{|a|,\alpha,\nu} = \nu\left[280|a|^{10} - 680|a|^8 + 592|a|^6 - 216|a|^4 + 36|a|^2 - 2\right]$$
$$- \alpha\left[140|a|^9 - 270|a|^7 + 176|a|^5 - 42|a|^3 + 6|a|\right] \quad (91)$$

We can check the expressions obtained from these formulas by using the special case $|a|=1$ and $\alpha=0$ for which

$$\langle x \rangle_t^{1,0,\nu} = 2\nu t \quad (92)$$

that we have obtained using equation 67.

These first 5 moments are plotted for $\nu = \frac{1}{2}$ and $\alpha = 0$ in figure 5. The time values of the graphs can be easily determined by noting for $\nu = \frac{1}{2}$ the terminating value $\langle x \rangle_t^{1,0,\frac{1}{2}} = t$ corresponds to the time.

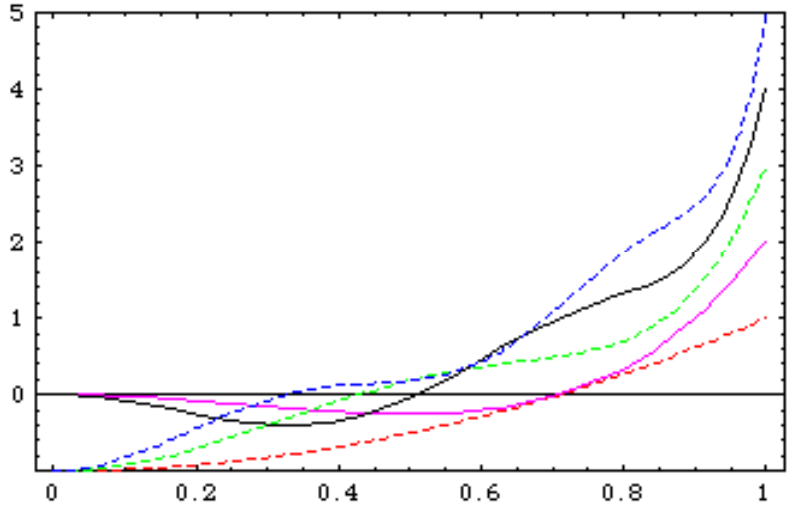

Figure 5. The first moments at early times

These moments coupled with the variance given in the next section are very revealing of the character of the quantum walk that corresponds to a particular value of $|a|$.

For small $|a|$ we note that these walks can be considered to oscillate with time between 0 and -1. The veracity of this statement is born witness to by the small value of the variance for small $|a|$ see figure 6. These oscillations become more damped with time as $|a|$ is increased up to a value of about 0.2. For large $|a|$ the walks increase linearly with time. The intermediate region is characterised by the largest variances as seen in figure 6.

## 7.3 Variance

The variance (*Var*) of the quantum walk can be evaluated from the expression [16]

$$Var(|a|,v,\alpha,t) = \langle x^2 \rangle_t^{|a|} - \left(\langle x \rangle_{t,v,\alpha}^{|a|}\right)^2. \qquad (88)$$

The normalised variances $Var(|a|,v,\alpha,t)/t^2$ are plotted for the times up to $t = 5$ and for $v = \frac{1}{2}$ and $\alpha = 0$ in figure 6. In interpreting these variances we note that the quantum walks are almost unimodal for these parameter values.

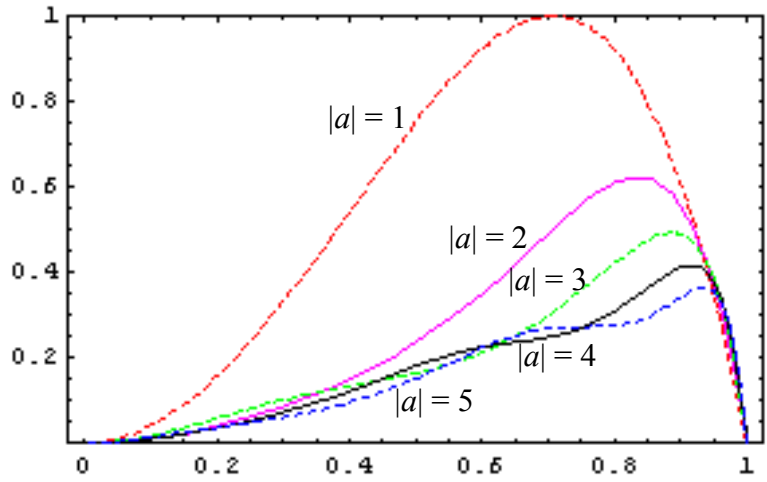

Figure 6. Normalised variances for $v = \frac{1}{2}$ and $\alpha = 0$

The dominant features of the variances are their zero values at the two extremes $|a| = 0$ and $|a| = 1$, and that their nonzero values between the end points vary continuously with the parameter $|a|$ and peak for large values with a rapid decrease to zero at $|a| = 1$.

## 8. The spectrum of quantum walks

An examination of the first and second moments and the variance indicates that the quantum walks have five regions of different behaviour related to the critical points extremes $|a| = 0$, $|a| = \frac{1}{\sqrt{2}}$ and $|a| = 1$. In the locality of the extremes $|a| = 0$ and $|a| = 1$ the variance is very small and the walk is typified by its dominant component. For small |a| the walks initially oscillate with time in a region bounded by $x = 0$ and $x$

= -1 with the size of the oscillations decreasing with time if |a| is nonzero. As |a| increases towards $|a|=\frac{1}{\sqrt{2}}$ the bounding breaks at earlier times and after this value the oscillations occur around a drift to larger values of x. The walks with $|a|\geq\frac{1}{\sqrt{2}}$ do not have any oscillations and are increasingly dominated by a drift to larger values of x with time that achieves its maximum at |a| = 1.

Thus Hadamard walks can be considered as typical of quantum walks with values of $||a|\geq\frac{1}{\sqrt{2}}$ but not for values smaller than this.

## 9. Conclusion

A general analysis of one dimensional quantum walks has been provided that gives new insights into their behaviour and tools for the modelling of experimental data.

In particular it has been shown that it is possible to express the general one dimensional quantum walk, in both the position and momentum space, in terms of simple analytic expressions. These expressions directly show that quantum walks depend on three effective real parameters, two that determine the symmetry of the walks and the third that controls the temporal evolution. The two that affect symmetry appear linearly in the expressions for the probability density of these walks making them easy to estimate from experimental data.

A new type of function was presented and partially analysed in terms of recursions and power series. This function was used to obtain the moments of the quantum walk as algebraic expressions of the walk parameters. These analytic expressions permitted an analysis of the walks that showed that the general second moments converge rapidly with time to a limiting form. They were then used to discuss the limited sense in which the Hadamard walk can be considered to be typical of the general one-dimensional quantum walk.

In the light of the strong current interest in quantum computing this analysis raises the open question of what insights for quantum algorithm development can be gleaned from general analytic models.

# Appendix A

The type II Chebyshev polynomials obey the recursion

$$U_{n+1}(x) = 2x U_n(x) - U_{n-1}(x). \qquad (A1)$$

Hence

$$U_{t+1}(|a|\cos(p)) = |a|(e^{ip} + e^{-ip})U_t(|a|\cos(p)) - U_{t-1}(|a|\cos(p)). \quad (A2)$$

The inverse Fourier transform of this relation is the recursion

$$\hat{U}_{t+1}(|a|:x) = |a|\hat{U}_t(|a|:x+1) + |a|\hat{U}_t(|a|:x-1) - \hat{U}_{t-1}(|a|:x). \qquad (A3)$$

Thus from equation 55 in the main text we obtain the recursion for the polynomials

$$P_k^{t+1}(|a|) = |a|P_{k+1}^t(|a|) + |a|P_{k-1}^t(|a|) - P_k^{t-1}(|a|). \qquad (A4)$$